\documentclass[a4paper,UKenglish,cleveref, autoref, thm-restate]{lipics-v2021}
\usepackage{style}

\title{Complete Bidirectional Typing for the Calculus of Inductive Constructions}

\titlerunning{Complete Bidirectional Typing for the Calculus of Inductive Constructions} %optional, please use if title is longer than one line

\author{Meven Lennon-Bertrand}{LS2N, Université de Nantes — Gallinette Project Team, Inria, France \and \url{http://www.meven.ac} }{meven.bertrand@univ-nantes.fr}{}{}%mandatory, please use full name; only 1 author per \author macro; first two parameters are mandatory, other parameters can be empty. Please provide at least the name of the affiliation and the country. The full address is optional

\authorrunning{M. Lennon-Bertrand} %mandatory. First: Use abbreviated first/middle names. Second (only in severe cases): Use first author plus 'et al.'

\Copyright{Meven Lennon-Bertrand} %mandatory, please use full first names. LIPIcs license is "CC-BY";  http://creativecommons.org/licenses/by/3.0/

 %mandatory: Please choose ACM 2012 classifications from https://dl.acm.org/ccs/ccs_flat.cfm 

 % mandatory; please add comma-separated list of keywords

\category{} %optional, e.g. invited paper

\relatedversion{} %optional, e.g. full version hosted on arXiv, HAL, or other repository/website
\supplementdetails[subcategory = {Formalization}]{Software}{https://github.com/MevenBertrand/metacoq/tree/itp-artefact}
\acknowledgements{Many thanks to Matthieu Sozeau for his help with MetaCoq and its nasty inductives, and to Chantal Keller, Nicolas Tabareau and the anonymous reviewers for their helpful comments on earlier versions of this article.}%TODO optional

\EventEditors{John Q. Open and Joan R. Access}
\EventNoEds{2}
\EventLongTitle{42nd Conference on Very Important Topics (CVIT 2016)}
\EventShortTitle{CVIT 2016}
\EventAcronym{CVIT}
\EventYear{2016}
\EventDate{December 24--27, 2016}
\EventLocation{Little Whinging, United Kingdom}
\EventLogo{}
\SeriesVolume{42}
\ArticleNo{23}
\begin{document}
\nolinenumbers

\maketitle

\begin{abstract}
  This article presents a bidirectional type system for the Calculus of Inductive Constructions (CIC). It introduces a new judgement intermediate between the usual inference and checking, dubbed constrained inference, to handle the presence of computation in types. The key property of the system is its completeness with respect to the usual undirected one, which has been formally proven in Coq as a part of the MetaCoq project. Although it plays an important role in an ongoing completeness proof for a realistic typing algorithm, the interest of bidirectionality is wider, as it gives insights and structure when trying to prove properties on CIC or design variations and extensions. In particular, we put forward constrained inference, an intermediate between the usual inference and checking judgements, to handle the presence of computation in types.
\end{abstract}
  
\section{Introduction}

  In logical programming, an important characteristic of judgements is the \emph{mode} of the objects involved, i.e., which ones are considered inputs or outputs. When examining this distinction for a typing judgement $\Gamma \vdash t : T$, both the term $t$ under inspection and the context $\Gamma$ of this inspection are usually known, they are thus inputs (although some depart from this, see \cite{Jim1996}). The mode of the type $T$, however, may vary: should it be inferred based upon $\Gamma$ and $t$, or do we merely want to check whether $t$ conforms to a given $T$? Both are sensible approaches, and in fact typing algorithms for complex type systems usually alternate between them during the inspection of a single term/program. The bidirectional approach makes this difference between modes explicit, by decomposing undirected\footnote{We call anything related to the $\Gamma \vdash t : T$ judgement undirected by contrast with the bidirectional typing.} typing $\Gamma \vdash t : T$ into two separate judgments $\Gamma \vdash t \inferty T$ (inference) and $\Gamma \vdash t \checkty T$ (checking)\footnote{We chose $\inferty$ and $\checkty$ rather than the more usual $\Rightarrow$ and $\Leftarrow$ to avoid confusion with implication on paper, and with the Coq notation for functions in the development.}, that differ only by modes. This decomposition allows theoretical work on practical typing algorithms, but also gives a finer grained structure to typing derivations, which can be of purely theoretical interest even without any algorithm in sight.

  Although this seems appealing, and despite advocacy by McBride \cite{McBride2018,McBride2019} to adopt this approach when designing type systems, most of the theoretical work on dependent typing to this day remains undirected. Others have described on paper bidirectional algorithms for dependent types, starting with Coquand \cite{Coquand1996} and continuing with Norell \cite{Norell2007} or Abel \cite{Abel2017}. However, all of these consider unannotated $\lambda$-abstractions, and use bidirectional typing as a way to remedy this lack of annotations. This is sensible when looking for lightness of the input syntax, but poses an inherent completeness problem, as a term like $(\l x . x)~0$ does not type-check against type $\nat$ in those systems. Very few have considered the case of annotated abstractions, apart from Asperti and the Matita team \cite{Asperti2012}, who however concentrate on specific problems pertaining to unification and implementation of the Matita elaborator, without giving a general bidirectional framework. They also do not consider the problem of completeness with respect to a given undirected system, as it would fail in their setting due to the undecidability of higher order unification.

  Thus, we wish to fill a gap in the literature, by describing a bidirectional type system that is complete with respect to the (undirected) Calculus of Inductive Constructions (CIC). By completeness, we mean that any term that is typable in the undirected system should also infer a type in the bidirectional one. This feature is very desirable when implementing kernels for proof assistants, whose algorithms should correspond to their undirected specification, never missing any typable term. The bidirectional systems we describe thus form intermediate steps between actual algorithms and undirected type systems. This step has proven useful in an ongoing completeness proof of MetaCoq's \cite{Sozeau2020a} type-checking algorithm\footnote{A completeness bug in that algorithm – also present in the Coq kernel – has already been found, see \cref{sec:to-pcuic} for details.}: rather than proving the algorithm complete directly, the idea is to prove it equivalent to the bidirectional type system, separating the implementation problems from the ones regarding the bidirectional structure.
  
  But the interest of having a bidirectional type system equivalent to the undirected one is not limited to the link with algorithms, it is also purely theoretical. First, the structure of a bidirectional derivation is more constrained than that of an undirected one, especially regarding the uses of computation. This finer structure can make proofs easier, while the equivalence ensures that they can be transported to the undirected world. For instance, in a setting with cumulativity/subtyping, the inferred type for a term $t$ should by construction be smaller than any other types against which $t$ checks. This provides an easy proof of the existence of principal types in the undirected system. The bidirectional structure also provides a better base for new type systems. This was actually the starting point for this investigation: in \cite{LennonBertrand2020}, we quickly describe a bidirectional variant of CIC, as the usual undirected CIC is unfit for the gradual extension we envision due to the too high flexibility of a free-standing conversion rule. This is the system we wish to thoroughly describe and investigate here.

  \subparagraph{Outline}
  We start by giving in \cref{sec:ccomega} a general roadmap in the simple setting of pure type systems, including the introduction of a constrained inference judgement that had not been clearly singled out in previous works.
  With the ideas set clear, we go on to the real thing: a bidirectional type system proven equivalent to the Predicative Calculus of Cumulative Inductive Constructions – PCUIC, the extension of CIC nowadays at the heart of Coq. This equivalence has been formalised on top of MetaCoq \cite{Sozeau2020}\footnote{A version frozen as described in this article is available in the following git branch: \url{https://github.com/MevenBertrand/metacoq/tree/itp-artefact}.}
  We next turn back to less technical considerations, as we believe that the bidirectional structure is of general theoretical interest. \Cref{sec:beyond} thus describes the value of basing type systems on the bidirectional system directly rather than on the equivalent undirected one. Finally \cref{sec:related} investigates related work, and in particular tries and identify the implicit presence of constrained inference in various earlier articles, showing how making it explicit clarifies those.

\section{Warming up with \CCo{}}
  \label{sec:ccomega}

\subsection{Undirected \CCo{}}

  As a starting point, let us consider the system \CCo{}. It is the backbone of CIC, and we can already illustrate most of our methodology on it. \CCo{} belongs to the wider class of pure type systems (PTS), that has been thoroughly studied and described, see for instance \cite{Barendregt1992}. Since there are many presentational variations, let us first give a precise account of our conventions. \defemph{Terms} in \CCo{} are given by the grammar
    \begin{mathpar}
    t ::= x \mid \uni[i] \mid \P x : t . t \mid \l x : t . t \mid t~t
    \end{mathpar}
  where the letter $x$ denotes a variable (so will letters $y$ and $z$), and the letter $i$ is an integer (we will also use letters $j$, $k$ and $l$ for those). All other Latin letters will be used for terms, with the upper-case ones used to suggest the corresponding terms should be though of as types — although this is not a syntactical separation. We abbreviate $\P x : A . B$ by $A \to B$ when $B$ does not depend on $x$, as is customary. On those terms, \defemph{reduction} $\red$ is defined as the least congruence such that $(\l x : T . t)~u \red \subs{t}{u}{x}$, where $\subs{t}{u}{x}$ denotes \defemph{substitution}.
  \defemph{Conversion} $\conv$ is the symmetric, reflexive, transitive closure of reduction. Finally, \defemph{contexts} are lists of variable declarations $x : t$ and are denoted using capital Greek letters. We write $\cdot$ for the empty list, $\Gamma, x : T$ for concatenation, and $(x : T) \in \Gamma$ if $(x : T)$ appears in $\Gamma$. Combining those, we can define \defemph{typing} $\Gamma \vdash t : T$ as in \cref{fig:pts-typing}, where $\max{i}{j}$ denotes the maximum of $i$ and $j$.
  We say a context $\Gamma$ is \defemph{well-formed} if $\vdash \Gamma$, a type $T$ is \defemph{well-formed} in a context $\Gamma$ if there exists $i$ such that $\Gamma \vdash T : \uni[i]$, and a term is \defemph{well-formed} in a context $\Gamma$ if there exists $T$ such that $\Gamma \vdash t : T$. We also use \defemph{well-typed} for the latter, and leave the context implicit for the last two when it is clear from context.
  These rules differ from more usual PTS presentations such as \cite{Barendregt1992} on the \nameref{infrule:pts-undir-var} and \nameref{infrule:pts-undir-sort} rules so as to avoid general weakening (which is however admissible) and single out the context well-formedness judgment. Premises are not minimal in order to provide more generous inductive hypotheses when doing proofs by induction on derivations. However, this presentation can easily be seen to be equivalent to that of \cite{Barendregt1992}.

  \begin{figure}
    \begin{mathpar}
      \jformlow{$\vdash \Gamma$}
      \inferrule{ }{\vdash \cdot}[Empty] \label{infrule:pts-undir-empty} \and
      \inferrule{\vdash \Gamma \\ \Gamma \vdash A : \uni[i]}{\vdash \Gamma, x : A}[Ext] \label{infrule:pts-undir-ext} \\
      \jform{$\Gamma \vdash t : T$}
      \inferrule{\vdash \Gamma}{\Gamma \vdash \uni[i] : \uni[i+1]}[Sort] \label{infrule:pts-undir-sort} \and
      \inferrule{\vdash \Gamma \\ (x : A \in \Gamma)}{\Gamma \vdash x : A}[Var] \label{infrule:pts-undir-var} \and
      %\inferrule{\Gamma \vdash x : A \\ \Gamma \vdash B : \uni[i]}{\Gamma, y : B \vdash x : A}[Weak] \label{infrule:pts-undir-weak} \and
      \inferrule{\Gamma \vdash A : \uni[i] \\ \Gamma, x : A \vdash B : \uni[j]}{\Gamma \vdash \P x : A . B : \uni[\max{i}{j}]}[Prod] \label{infrule:pts-undir-prod} \and
      \inferrule{\Gamma \vdash \P x : A . B : \uni[i] \\ \Gamma, x : A \vdash t : B}{\Gamma \vdash \l x : A . t : \P x : A . B}[Abs] \label{infrule:pts-undir-abs} \and
      \inferrule{\Gamma \vdash t : \P x : A . B \\ \Gamma \vdash u : A}{\Gamma \vdash t~u : \subs{B}{u}{x}}[App] \label{infrule:pts-undir-app} \and
      \inferrule{\Gamma \vdash t : A \\ \Gamma \vdash B : \uni[i] \\ A \conv B}{\Gamma \vdash t : B}[Conv] \label{infrule:pts-undir-conv}
    \end{mathpar}

    \caption{Undirected typing for \CCo{}}
    \label{fig:pts-typing}
  \end{figure}

  As any PTS, \CCo{} has many desirable properties. We summarize the ones we rely on here. Detailed proofs in the context of PTS can be found in \cite{Barendregt1992}, and formalisation of the corresponding properties for PCUIC are an important part of MetaCoq.

  \begin{proposition}[Properties of \CCo{}]
    \label{prop:ccomega}
    The type system \CCo{} as just described enjoys the following properties:
    \begin{description}
      \item[Confluence] Reduction $\red$ is confluent. As a direct consequence, two terms are convertible just when they have a common reduct: $t \conv u$ if and only if there exists $t'$ such that $t \rtred t'$ and $u \rtred t'$.
      \item[Transitivity] Conversion is transitive.
      \item[Subject reduction] If $\Gamma \vdash t : T$ and $t \red t'$ then $\Gamma \vdash t' : T$.
      \item[Validity] If $\Gamma \vdash t : T$ then $T$ is well-formed, e.g.\ there exists some $i$ such that $\Gamma \vdash T : \uni[i]$.
    \end{description}
  \end{proposition}

\subsection{Turning \CCo{} Bidirectional}
  \label{sec:ccomega-bidir}

  \subparagraph{McBride’s discipline}
  To design our bidirectional type system, we follow a discipline exposed by McBride \cite{McBride2018,McBride2019}. The central point is to distinguish in a judgment between the subject, whose well-formedness is under scrutiny, from inputs, whose well-formedness is a condition for the judgment to behave well, and outputs, whose well-formedness is a consequence of the judgment. For instance, in inference $\Gamma \vdash t \inferty T$, the subject is $t$, $\Gamma$ is an input and $T$ is an output. This means that one should consider whether $\Gamma \vdash t \inferty T$ only in cases where $\vdash \Gamma$ is already known, and if the judgment is derivable it should be possible to conclude that both $t$ and $T$ are well-formed. All inference rules are to preserve this invariant. This means that inputs to a premise should be well-formed whenever the inputs to the conclusion and outputs and subjects of previous premises are. Similarly the outputs of the conclusion should be well-formed if the inputs of the conclusion and the subjects and outputs of the premises are assumed to be so.
  
  This distinction also applies to the computation-related judgments, although those have no subject. For conversion $T \conv T'$ both $T$ and $T'$ are inputs, and thus should be known to be well-formed beforehand. For reduction $T \rtred T'$, $T$ is an input and $T'$ is an output, so only $T$ needs to be well-formed, with the subject reduction property of the system ensuring that the output $T'$ is also well-formed.
  %\mlb{Would it make sense to use colors in the figure to emphasize mode?}

  \begin{figure}
    \begin{mathpar}
      % \jform{$\vdash \Gamma \checkty$} \\
      % \inferrule{ }{\vdash \cdot \checkty}[Empty] \label{infrule:pts-empty} \and
      % \inferrule{
      %   \vdash \Gamma \checkty \\
      %   \Gamma \vdash T \pinferty{\sorts} s}
      % {\vdash \Gamma, x : T \checkty}[Concat] \label{infrule:pts-concat} \\
    \jform{Inference: $\Gamma \vdash t \inferty T$}
      \inferrule{ }{\Gamma \vdash \uni[i] \inferty \uni[i+1]}[Sort] \label{infrule:pts-sort} \and
      \inferrule{(x : T) \in \Gamma}{\Gamma \vdash x \inferty T}[Var] \label{infrule:pts-var} \and
      \inferrule{\Gamma \vdash A \pinferty{\sorts} \uni[i] \\ \Gamma, x : A \vdash B \pinferty{\sorts} \uni[j]}
        {\Gamma \vdash  \P x : A . B \inferty \uni[\max{i}{j}]}[Prod] \label{infrule:pts-prod} \\ 
      \inferrule{\Gamma \vdash A \pinferty{\sorts} \uni[i] \\ \Gamma, x : A \vdash t \inferty B}
        {\Gamma \vdash \l x : A . t \inferty  \P x : A . B}[Abs] \label{infrule:pts-abs} \and
      \inferrule{\Gamma \vdash t \pinferty{\Pi}  \P x : A . B \\ \Gamma \vdash u \checkty A}
        {\Gamma \vdash t~u \inferty \subs{B}{u}{x}}[App] \label{infrule:pts-app} \\
    \jformlow{Checking: $\Gamma \vdash t \checkty T$} 
    \inferrule{\Gamma \vdash t \inferty T' \\ T' \conv T}
      {\Gamma \vdash t \checkty T}[Check] \label{infrule:pts-check} \\
    \jform{Constrained inference: $\Gamma \vdash t \pinferty{h} T$}
    \inferrule{\Gamma \vdash t \inferty T \\ T \rtred \uni[i]}
      {\Gamma \vdash t \pinferty{\sorts} \uni[i]}[Sort-Inf] \label{infrule:pts-sort-inf} \and
    \inferrule{\Gamma \vdash t \inferty T \\ T \rtred \P x : A. B}
      {\Gamma \vdash t \pinferty{\P} \P x : A . B}[Prod-Inf] \label{infrule:pts-prod-inf}
  \end{mathpar}
  
  \caption{Bidirectional typing for \CCo{}}
  \label{fig:bidir-pts}
  \end{figure}

  \subparagraph{Constrained inference}
  Beyond the already described inference and checking judgements another one appears in the bidirectional typing rules of \cref{fig:bidir-pts}: \defemph{constrained inference}, written $\Gamma \vdash t \pinferty{h} T$, where $h$ is either $\Pi$ or $\sorts$  – and will be extended once we introduce more type formers. Constrained inference is a judgement – or, rather, a family of judgements indexed by $h$ – with the exact same modes as inference, but where the type output is not completely free. Rather, as the name suggests, a constraint is imposed on it, namely that its head constructor can only be the corresponding element of $h$. This is needed to handle the behaviour absent in simple types that some terms might not have a desired type “on the nose”, as exemplified by the first premise $\Gamma \vdash t \pinferty{\Pi} \P x : A . B$ of the \nameref{infrule:pts-app} rule for $t~u$. Indeed, it would be too much to ask $t$ to directly infer a $\Pi$-type, as some reduction might be needed on $T$ to uncover this $\Pi$. Checking also cannot be used, because the domain and codomain of the tentative $\Pi$-type are not known at that point: they are to be inferred from $t$.

  \subparagraph{Structural rules}
  To transform the rules of \cref{fig:pts-typing} to those of \cref{fig:bidir-pts}, we start by recalling that we wish to obtain a complete bidirectional type system. Therefore any term should infer a type, and thus all structural rules (i.e.\ all rules where the subject of the conclusion starts with a term constructor) should give rise to an inference rule. It thus remains to choose the judgements for the premises, which amounts to determining their modes. If a term in a premise appears as input in the conclusion or output of a previous premise, then it can be considered an input, otherwise it must be an output. Moreover, if a type output is unconstrained, then inference can be used, otherwise we must resort to constrained inference.
  
  This applies straightforwardly to most rules but the PTS-style \nameref{infrule:pts-undir-abs}. Indeed, neither $\Gamma \vdash \P x : A . B : \uni[i]$ nor $\Gamma, x : A \vdash t : B$ can be taken as the first bidirectional premise: the first one because $B$ is not known from inputs to the conclusion, and the second because context $\Gamma, x : A$ is not known to be well-formed from the conclusion. For general PTS, this is quite problematic, as demonstrated by Pollack \cite{Pollack1992}. For \CCo{}, however, the solution is simple. Replacing $\Gamma \vdash \P x : A . B : \uni[i]$ by the equivalent $\Gamma \vdash A : \uni[j]$ and $\Gamma, x : A \vdash B : \uni[j']$, the former can become the first premise, ensuring that type inference for $t$ is done in a well-formed context, and the latter can be dropped based upon the invariant that outputs – here the type $B$ inferred for $t$ — are well-formed.

  Similarly, as the context is always supposed to be well-formed as an input to the conclusion, it is not useful to re-check it, and thus the premise of \nameref{infrule:pts-undir-sort} and \nameref{infrule:pts-undir-var} can be safely dropped. This is in line with implementations, where the context is not re-checked at leaves of a derivation tree, with performance issues in mind. The well-formedness invariants then ensure that any derivation starting with the (well-formed) empty context will only handle well-formed contexts.

  \subparagraph{Computation rules}
  We are now left with the non-structural conversion rule. As we observed, there are two possible modes for computation: if both sides are inputs, conversion can be used, but if only one is known one must resort to reduction, and the other side becomes an output instead. Rule \nameref{infrule:pts-check} corresponds to the first case, while rules \nameref{infrule:pts-prod-inf} and \nameref{infrule:pts-sort-inf} both are in the second case. This difference in turn introduces the need to separate between checking, that calls for the first rule, and constrained inference, that requires the others.

  The need to split the conversion rule into a (weak-head) reduction and conversion subroutine depending on the mode is known to the implementors of proof assistants \cite{Abel2011}. However, we wish to de-emphasize the role devoted specifically to reduction in the description of those algorithms, instead putting constrained inference forward. Indeed, reduction is only a means to determine whether a certain term fits into a certain category of types. In the setting of \CCo{}, there is mainly one way to do so, which is to reduce its type until its head constructor is exposed.
  However, as soon as conversion is extended, for instance with unification \cite{Asperti2012}, coercions \cite{Asperti2012,Sozeau2007} or graduality \cite{LennonBertrand2020}, reduction is not enough any more. Singling out constrained inference then makes the required modification to the typing rules clearer. We come back to this more in length in \cref{sec:rel-constrained}.

\subsection{Properties}

  Let us now state the two main properties relating the bidirectional system to the undirected one: it is both correct (terms typable in the bidirectional system are typable in the undirected system) and complete (all terms typable in the undirected system are also typable in the bidirectional system).

\subsubsection{Correctness}

  A bidirectional derivation can be seen as a refinement of an undirected derivation. Indeed, the bidirectional structure can be erased – replacing each bidirectional rule with the corresponding undirected rule – to obtain an undirected derivation, but for missing sub-derivations, which can be retrieved using the invariants on well-formedness of inputs and outputs. Thus, we get the following correctness theorem – note how McBride’s discipline manifests as well-formedness hypothesis on inputs.

  \begin{theorem}[Correctness of bidirectional typing for \CCo{}]
    \label{thm:corr-ccomega}
    If $\Gamma$ is well-formed and $\Gamma \vdash t \inferty T$ or $\Gamma \vdash t \pinferty{h} T$ then $\Gamma \vdash t : T$. If $\Gamma$ and $T$ are well-formed and $\Gamma \vdash t \checkty T$ then $\Gamma \vdash t : T$. 
  \end{theorem}
  
  \begin{proof}
    The proof is by mutual induction on the bidirectional typing derivation.

    Each rule of the bidirectional system can be replaced by the corresponding rule of the undirected system, with all three \nameref{infrule:pts-check}, \nameref{infrule:pts-prod-inf} and \nameref{infrule:pts-sort-inf} replaced by \nameref{infrule:pts-undir-conv}, \nameref{infrule:pts-abs} using an extra \nameref{infrule:pts-undir-prod} rule. In all cases, the induction hypothesis can be used on sub-derivations of the bidirectional judgment because the context extensions and checking are done with types that are known to be well-formed by induction hypothesis on previous premises and validity.

    Some sub-derivations of the undirected rules that have no counterpart in the bidirectional ones are however missing. In rules \nameref{infrule:pts-sort} and \nameref{infrule:pts-var} the hypothesis that $\vdash \Gamma$ is enough to get the required premise. For rule \nameref{infrule:pts-check}, the well-formedness hypothesis on the type is needed to get the second premise of rule \nameref{infrule:pts-undir-conv}. As for \nameref{infrule:pts-prod-inf} and \nameref{infrule:pts-sort-inf}, that second premise is obtained using the induction hypothesis, validity and subject reduction. Finally, the missing premise on the codomain of the product type in rule \nameref{infrule:pts-abs} is obtained by validity. 
    
    Uses of validity could alternatively be handled by strengthening the theorem to incorporate the well-formedness of types when they are outputs.

  \end{proof}

\subsubsection{Completeness}

  Let us now state the most important property of our bidirectional system: it does not miss any undirected derivation.

  \begin{theorem}[Completeness of bidirectional typing for \CCo{}]
    \label{thm:compl-ccomega}
    If $\Gamma \vdash t : T$ then there exists $T'$ such that $\Gamma \vdash t \inferty T'$ and $T' \conv T$.
  \end{theorem}

  \begin{proof}
    The proof is by induction on the undirected typing derivation.
    
    Rules \nameref{infrule:pts-undir-sort} and \nameref{infrule:pts-undir-var} are base cases, and can be replaced by the corresponding rules in the bidirectional world. Rule \nameref{infrule:pts-undir-conv} is a direct consequence of the induction hypothesis on its first premise, together with transitivity of conversion.
    
    For rule \nameref{infrule:pts-undir-prod}, we need the intermediate lemma that if $T$ is a term such that $T \conv \uni[i]$, then also $T \rtred \uni[i]$. This is a consequence of confluence of reduction. In turn, it implies that if $\Gamma \vdash t \inferty T$ and $T \conv \uni[i]$ then $\Gamma \vdash t \pinferty{\sorts} \uni[i]$, and is enough to conclude for that rule.
    
    In rule \nameref{infrule:pts-undir-abs}, the induction hypothesis gives $\Gamma \vdash \P x : A . B \inferty T$ for some $T$, and an inversion on this gives $\Gamma \vdash A \pinferty{\uni} \uni[i]$ for some $i$. Combined with the second induction hypothesis, we get $\Gamma \vdash \l x : A . t \inferty \P x : A . B'$ for some $B'$ such that $B \conv B'$, and thus $\P x : A . B \conv \P x : A . B'$ as desired.

    We are finally left with the \nameref{infrule:pts-undir-app} rule. We know that $\Gamma \vdash t \inferty T$ with $T \conv \P x : A . B$. Confluence then implies that $T \rtred \P x : A' . B'$ for some $A'$ and $B'$ such that $A \conv A'$ and $B \conv B'$. Thus $\Gamma \vdash t \pinferty{\Pi} \P x : A' . B'$. But by induction hypothesis we also know that $\Gamma \vdash u \inferty A''$ with $A'' \conv A$ and so by transitivity of conversion $\Gamma \vdash u \checkty A'$. We can thus apply \nameref{infrule:pts-app} to conclude.

  \end{proof}

  Contrarily to correctness, which kept a similar derivation structure, completeness is of a different nature. Because in bidirectional derivations the conversion rules are much less liberal than in undirected derivations, the crux of the proof is to ensure that conversions can be permuted with structural rules, in order to concentrate them in the places where they are authorized in the bidirectional derivation. In a way, composing completeness with conversion gives a kind of normalization procedure that produces a canonical undirected derivation by pushing all conversions down as much as possible.

\subsubsection{Reduction strategies}

  The judgements of \cref{fig:bidir-pts} are syntax-directed, in the sense that there is always at most one rule that can be used to derive a certain typing judgements. But with the rules as given there is still some indeterminacy. Indeed when appealing to reduction no strategy is fixed, thus two different reducts give different uses of the rule, resulting in different inferred types – although those are still convertible. However, a reduction strategy can be imposed to completely eliminate indeterminacy in typing, leading to the following.
  
  \begin{proposition}[Reduction strategy]
    \label{prop:red-strat}

    If $\rtred$ is replaced by weak-head reduction in rules \nameref{infrule:pts-sort-inf} and \nameref{infrule:pts-prod-inf}, then given a well-formed context $\Gamma$ and a term $t$ there is at most one derivation of $\Gamma \vdash t \inferty T$ and $\Gamma \vdash t \pinferty{h} T$, and so in particular such a $T$ is unique. Similarly, given well-formed $\Gamma$ and $T$ and a term $t$ there is at most one derivation of $\Gamma \vdash t \checkty T$.
    Moreover, the existence of those derivations is decidable.
  \end{proposition}

  The algorithm for deciding the existence of the derivations is straightforward from the bidirectional rules, it amounts to structural recursion on the subject.

\section{From \CCo{} to PCUIC}
  \label{sec:to-pcuic}
  
  \CCo{} is already a powerful system, but today’s proof assistants rely on much more complex features. There are two main areas of differences between \CCo{} and the Predicative Calculus of Cumulative Inductive Constructions (PCUIC), the type theory nowadays behind the Coq proof assistant. Adapting to those is a good stress test for the bidirectional approach: seamlessly doing so is a good sign that the general methodology we presented is robust.

  The first area of difference are the universes. While on paper those are simply integer, to handle typical ambiguity and polymorphic (co)-inductive types, PCUIC uses algebraic universes, containing level variables, algebraic $\vee$ and $+1$ operators, and a special level for the sort Prop. Moreover, those universes are cumulative, that is they behave as if smaller universes were included in larger ones. The precise handling of the algebraic universes is abstracted away in MetaCoq, and quite similar in the directed and undirected systems, so it did not prove too difficult to handle. Cumulativity, however, introduces some not-so-small differences with the previous presentation, so we spend some time on it in \cref{sec:pcuic-cumul}.

  The second is the addition of new base type and term constructors. We describe the treatment of inductive types in \cref{sec:pcuic-indu}. Co-inductive types and records behave very similarly to inductive types at the level of typing, so we do not dwell on them. The difference lies mainly at the level of reduction/conversion, but as our type system treats those as black boxes the differences have a negligible impact.
  
  The interplay between those two areas is quite subtle, and we were able to uncover an incompleteness bug in the current kernel of Coq regarding pattern-matching of cumulative inductive types. This bug had gone unnoticed until our formalisation, but was causing subject reduction failures in some corner cases\footnote{The precise issue in the kernel is described in this git issue: \url{https://github.com/coq/coq/issues/13495}.}. We try and give an intuition of the problem in \cref{sec:pcuic-indu}.

\subsection{Cumulativity}
\label{sec:pcuic-cumul}

  PCUIC incorporates a limited form of subtyping, corresponding to the intuition that smaller universes are included in larger ones. Conversion $\conv$ is therefore replaced by \defemph{cumulativity} $\cumul$, the main difference being that the constraint on universes is relaxed. For conversions $\uni[i] \conv \uni[j]$ only when $i = j$, but for cumulativity $\uni[i] \cumul \uni[j]$ whenever $i \leq j$ – and this extends by congruence through most constructors. The conversion rule is accordingly replaced by the following cumulativity rule
  \begin{mathpar}
    \inferrule{\Gamma \vdash t : A \\ \Gamma \vdash B : \uni[i] \\ A \cumul B }{\Gamma \vdash t : B}[Cumul] \label{infrule:pcuic-undir-cumul}
  \end{mathpar}
  This reflects the view that universes $\uni[i]$ should be included one in the next when going up in the hierarchy. In \CCo{}, all types for a given term $t$ in a fixed context $\Gamma$ are equally good, as they are all convertible. This is not the case any more in presence of cumulativity, as we can have $T \cumul T'$ but not $T \conv T'$. Of particular interest are principal types, defined as follows.

  \begin{definition}[Principal type]
    The term $T$ is called a \defemph{principal type} for term $t$ in context $\Gamma$ if it is a least type for $t$ in $\Gamma$, that is if $\Gamma \vdash t : T$ and for any $T'$ such that $\Gamma \vdash t : T'$ we have $T \cumul T'$.
  \end{definition}

  The existence of such principal types is no so easy to prove directly but quite useful, as they are in a sense the best types for any terms. Indeed, if $T$ is a principal type for $t$ in $\Gamma$ and $T'$ is any other type for $t$, the \nameref{infrule:pcuic-undir-cumul} rule can be used to deduce $\Gamma \vdash t : T'$ from $\Gamma \vdash t : T$, which in general is not the case if $T$ is not principal. Similarly, if $T$ and $T'$ are two types for a term $t$, then they are not directly related, but the existence of principal types ensures that there exists some $T''$ that is a type for $t$ and such that $T \cumul T'$ and $T \cumul T''$, indirectly relating $T'$ and $T''$.

  Reflecting this modification in the bidirectional system of course calls for an update to the computation rules. The change to the \nameref{infrule:pts-check} rule is direct: simply replace conversion with cumulativity.
  \begin{mathpar}
    \inferrule{\Gamma \vdash t \inferty A \\ A \cumul B }{\Gamma \vdash t \checkty B}[Cumul] \label{infrule:pcuic-cumul}
  \end{mathpar}
  As to the constrained inference rules, there is no need to modify them. Intuitively, this is because there is no reason to degrade a type to a larger one when it is not needed. We only resort to cumulativity when it is forced by a given input. In that setting, completeness becomes the following:

  \begin{theorem}[Completeness with cumulativity]
    \label{thm:comp-cumul}
    If $\Gamma \vdash t : T$ using rules of \cref{fig:pts-typing} replacing \nameref{infrule:pts-undir-conv} with \nameref{infrule:pcuic-undir-cumul}, then $\Gamma \vdash t \inferty T'$ is derivable with rules of \cref{fig:bidir-pts} replacing \nameref{infrule:pts-check} with \nameref{infrule:pcuic-cumul} for some $T'$ such that $T' \cumul T$.
  \end{theorem}

  In that setting, even without fixing a reduction strategy as in \cref{prop:red-strat}, there is a weaker uniqueness property for inferred types.

  \begin{proposition}[Uniqueness of inferred type]
    \label{prop:unique-inf}
    If $\Gamma$ is well-formed, $\Gamma \vdash t \inferty T$ and $\Gamma \vdash t \inferty T'$ then $T \conv T'$. Similarly if $\Gamma$ is well-formed, $\Gamma \vdash t \pinferty{h} T$ and $\Gamma \vdash t \pinferty{h} T'$ then $T \conv T'$.
  \end{proposition}

  \begin{proof}
    Mutual induction on the first derivation. It is key that constrained inference rules only reduce a type, so that the type in the conclusion is convertible to the type in the premise, rather than merely related by cumulativity.
  \end{proof}
  
  In particular, combining those two properties with a correctness property akin to \cref{thm:corr-ccomega}, we can prove that any inferred type is principal, and so that they both exist and are computable since the bidirectional judgement can still be turned into an algorithm in the spirit of \cref{prop:red-strat}.

  \begin{proposition}[Principal types]
    \label{prop:princ-types}
    If $\Gamma$ is well-formed and $\Gamma \vdash t \inferty T$ then $T$ is a principal type for $t$ in $\Gamma$.
  \end{proposition}

  \begin{proof}
    If $\Gamma \vdash t : T'$, then by completeness there exists some $T''$ such that $\Gamma \vdash t \inferty T''$ and moreover $T'' \cumul T'$. But by uniqueness $T \conv T'' \cumul T'$ and thus $T \cumul T'$, and $T$ is indeed a principal type for $t$ in $\Gamma$.
  \end{proof}

  Reasoning on the bidirectional derivation thus makes proofs easier, while the correctness and completeness properties ensure they can be carried to the undirected system. Another way to understand this is that seeing completeness followed by correction as a normalization procedure on derivations, the produced canonical derivation is more structured and thus more amenable to proofs. Here for instance the uniqueness of the inferred type translates to the existence of principal types via correctness, and the normalization of the derivations optimizes it to derive a principal type.

\subsection{Inductive Types}
\label{sec:pcuic-indu}

  \begin{figure}[th]
    
    \begin{mathpar}
  %     \inferrule{\vdash \Gamma}{\Gamma \vdash \nat : \uni[0]} \and
  %     \inferrule{\vdash \Gamma}{\Gamma \vdash \zero : \nat} \and
  %     \inferrule{\vdash \Gamma}{\Gamma \vdash \succ : \nat \to \nat} \and
  %     \inferrule{
  %       \Gamma, x : \nat \vdash P : \uni[i] \\
  %       \Gamma \vdash p_{\zero} : \subs{P}{\zero}{x} \\
  %       \Gamma, n : \nat, p : \subs{P}{n}{x} \vdash p_{\succ} : \subs{P}{\succ[n]}{x} \\
  %       \Gamma \vdash s : \nat }
  %       {\Gamma \vdash \indnat{s}{x.P}{p_{\zero}}{p_{\succ}} : \subs{P}{s}{x}} \\
  %     %
      \inferrule{\Gamma \vdash A : \uni[i] \\ \Gamma, x : A \vdash B : \uni[j]}{\Gamma \vdash \S x : A . B : \uni[\max{i}{j}]}[$\S$-type] \label{infrule:undir-sig-type} \and
      \inferrule{\Gamma \vdash A : \uni[i] \\ \Gamma, x : A \vdash B : \uni[j] \\
        \Gamma \vdash a : A \\ \Gamma \vdash b : \subs{B}{a}{x}}{\Gamma \vdash \pair{x}{A}{B}{a}{b} : \S x : A . B}[$\S$-cons] \label{infrule:undir-sig-cons} \and
      \inferrule{
        \Gamma, z : \S x : A . B \vdash P : \uni[i] \\
        \Gamma, x : A, y : B \vdash b : \subs{P}{(x,y)}{z} \\
        \Gamma \vdash s : \S x : A . B }
        {\Gamma \vdash \indS{z.P}{x.y.p}{s} : \subs{P}{s}{z}}[$\S$-rec] \label{infrule:undir-sig-rec} \\
  %     %
  %     \inferrule{\Gamma \vdash A : \uni[i] \\ \Gamma \vdash a : A \\ \Gamma \vdash a' : A}{\Gamma \vdash \eq{A}{a}{a'} : \uni[i]} \and
  %     \inferrule{\Gamma \vdash A : \uni[i] \\ \Gamma \vdash a : A}{\Gamma \vdash \refl{A}{a} : \eq{A}{a}{a}} \and
  %     \inferrule{
  %       \Gamma, x : A, y : \eq{A}{a}{x} \vdash P : \uni[i] \\
  %       \Gamma \vdash p : \subs{\subs{P}{\refl{A}{a}}{y}}{a}{x} \\
  %       \Gamma \vdash s : \eq{A}{a}{a'} }
  %       {\Gamma \vdash \indeq{s}{x.y.P}{p} : \subs{\subs{P}{s}{y}}{a'}{x}}
    \end{mathpar}

    \caption{Undirected sum type}
    \label{fig:undir-indu}
  \end{figure}

  \begin{figure}[th]
    
    \begin{mathpar}
      \jformlow{$\Gamma \vdash t \inferty T$}
      % \inferrule{ }{\Gamma \vdash \nat \inferty \uni[0]} \and
      % \inferrule{ }{\Gamma \vdash \zero \inferty \nat} \and
      % \inferrule{ }{\Gamma \vdash \succ \inferty \nat \to \nat} \and
      % \inferrule{
      %   \Gamma \vdash s \pinferty{\nat} \nat \\
      %   \Gamma, x : \nat \vdash P \pinferty{\uni} \uni[i] \\
      %   \Gamma \vdash p_{\zero} \checkty \subs{P}{\zero}{x} \\
      %   \Gamma, n : \nat, p : \subs{P}{n}{x} \vdash p_{\succ} \checkty \subs{P}{\succ[n]}{x} }
      %   {\Gamma \vdash \indnat{x.P}{p_{\zero}}{p_{\succ}{s}} \inferty \subs{P}{s}{x}} \\
      %
      \inferrule{\Gamma \vdash A \pinferty{\uni} \uni[i] \\ \Gamma, x : A \vdash B \pinferty{\uni} \uni[j]}{\Gamma \vdash \S x : A . B \inferty \uni[\max{i}{j}]}[$\S$-type] \label{infrule:sig-type} \and
      \inferrule{\Gamma \vdash A \pinferty{\uni} \uni[i] \\ \Gamma, x : A \vdash B \pinferty{\uni} \uni[j] \\
        \Gamma \vdash a \checkty A \\ \Gamma \vdash b \checkty \subs{B}{a}{x}}{\Gamma \vdash \pair{x}{A}{B}{a}{b} \inferty \S x : A . B}[$\S$-cons] \label{infrule:sig-cons} \and
      \inferrule{
        \Gamma \vdash s \pinferty{\S} \S x : A . B \\
        \Gamma, z : \S x : A . B \vdash P \pinferty{\uni} \uni[i] \\
        \Gamma, x : A, y : B \vdash b \checkty \subs{P}{(x,y)}{z} }
        {\Gamma \vdash \indS{z.P}{x.y.b}{s} \inferty \subs{P}{s}{z}}[$\S$-rec] \label{infrule:sig-rec} \\
      %
      % \inferrule{\Gamma \vdash A \pinferty{\uni} \uni[i] \\ \Gamma \vdash a \checkty A \\ \Gamma \vdash a' \checkty A}{\Gamma \vdash \eq{A}{a}{a'} \inferty \uni[i]} \and
      % \inferrule{\Gamma \vdash A \pinferty{\uni} \uni[i] \\ \Gamma \vdash a \checkty A}{\Gamma \vdash \refl{A}{a} \inferty \eq{A}{a}{a}} \and
      % \inferrule{
      %   \Gamma \vdash s \inferty \eq{A}{a}{a'} \\
      %   \Gamma, x : A, y : \eq{A}{a}{x} \vdash P \pinferty{\uni} \uni[i] \\
      %   \Gamma \vdash p \checkty \subs{\subs{P}{\eq{A}{a}{a}}{y}}{a}{x} }
      %   {\Gamma \vdash \indeq{x.y.P}{p}{s} \inferty \subs{\subs{P}{s}{y}}{a'}{x}} \\
      %
      \jformlow{$\Gamma \vdash t \pinferty{h} T$}
      % \inferrule{\Gamma \vdash t \inferty T \\ T \rtred \nat}{\Gamma \vdash t \pinferty{\nat} \nat} \and
      \inferrule{\Gamma \vdash t \inferty T \\ T \rtred \S x : A . B}{\Gamma \vdash t \pinferty{\S} \S x : A . B}[$\S$-Inf] \label{infrule:sig-inf} \and
      % \inferrule{\Gamma \vdash t \inferty T \\ T \rtred \eq{A}{a}{a'}}{\Gamma \vdash t \pinferty{\eqind} \eq{A}{a}{a'}}
    \end{mathpar}

    \caption{Bidirectional sum type}
    \label{fig:bidir-indu}
  \end{figure}

  \subparagraph{Sum type}

  Before we turn to the general case of inductive types of the formalisation, let us present a simple inductive type: dependent sums. The undirected rules are given in \cref{fig:undir-indu}, and are inspired from the theoretical presentation of such dependent sums, such at the one of the Homotopy Type Theory book \cite{UniFoundationsProgram2013}. In particular, we use the same convention to write $y.P$ when variable $y$ is bound in $P$. Note however that contrarily to \cite{UniFoundationsProgram2013}, some typing information is kept on the pair constructor. Exactly as for the abstraction, this is to be able to infer a unique, most general type in the bidirectional system. Indeed, without that information a pair $(a,b)$ could inhabit multiple types $\S x : A . B$ because there are potentially many incomparable types $B$ such that $\subs{B}{a}{x}$ is a type for $b$, as even if $\subs{B}{a}{x}$ and $\subs{B'}{a}{x}$ are convertible $B$ and $B'$ may be quite different, depending of which instances of $a$ in $\subs{B}{a}{x}$ are abstracted to $x$.

  To obtain the bidirectional rules of \cref{fig:bidir-indu}, first notice that all undirected rules are structural and must thus become inference rules if we want the resulting system to be complete, just as in \cref{sec:ccomega}. The question therefore is again to know which modes to choose for the premises. For \nameref{infrule:undir-sig-type} and \nameref{infrule:sig-cons} this is straightforward: when the type appears in the conclusion, use checking, otherwise (constrained) inference. The case of the destructors is somewhat more complex. Handling the subterms of the destructor in the order in which they usually appear (predicate, branches and finally scrutinee) is not possible, as the parameters of the inductive type are needed to construct the context in which the predicate is typed. However those parameters can be inferred from the scrutinee. Thus, a type for the scrutinee is first obtained using a new constrained inference judgment, forcing the inferred type to be a $\S$-type, but leaving its parameters free. Next, the obtained arguments can be used to construct the context to type the predicate. Finally, once the predicate is known to be well-formed, it can be used to type-check the branch. 
  
  \begin{figure}[th]
    
    \begin{mathpar}
      \jformlow{$\Gamma \vdash t \inferty T$}
      \inferrule{ }{\Gamma \vdash \nat \inferty \uni[0]} \and
      \inferrule{ }{\Gamma \vdash \zero \inferty \nat} \and
      \inferrule{\Gamma \vdash n \checkty \nat }{\Gamma \vdash \succ[n] \inferty \nat} \and
      \inferrule{
        \Gamma \vdash s \pinferty{\nat} \nat \\
        \Gamma, z : \nat \vdash P \pinferty{\uni} \uni[i] \\
        \Gamma \vdash b_{\zero} \checkty \subs{P}{\zero}{z} \\
        \Gamma, x : \nat, p : \subs{P}{x}{z} \vdash b_{\succ} \checkty \subs{P}{\succ[x]}{z} }
        {\Gamma \vdash \indnat{z.P}{b_{\zero}}{x.p.b_{\succ}}{s} \inferty \subs{P}{s}{z}} \\
      \inferrule{\Gamma \vdash A \pinferty{\uni} \uni[i] \\ \Gamma \vdash a \checkty A \\ \Gamma \vdash a' \checkty A}{\Gamma \vdash \eq{A}{a}{a'} \inferty \uni[i]} \and
      \inferrule{\Gamma \vdash A \pinferty{\uni} \uni[i] \\ \Gamma \vdash a \checkty A}{\Gamma \vdash \refl{A}{a} \inferty \eq{A}{a}{a}} \and
      \inferrule{
        \Gamma \vdash s \inferty \eq{A}{a}{a'} \\
        \Gamma, x : A, z : \eq{A}{a}{x} \vdash P \pinferty{\uni} \uni[i] \\
        \Gamma \vdash b \checkty \subs{\subs{P}{\eq{A}{a}{a}}{z}}{a}{x} }
        {\Gamma \vdash \indeq{x.z.P}{b}{s} \inferty \subs{\subs{P}{s}{z}}{a'}{x}} \\
      
      \jformlow{$\Gamma \vdash t \pinferty{h} T$}
      \inferrule{\Gamma \vdash t \inferty T \\ T \rtred \nat}{\Gamma \vdash t \pinferty{\nat} \nat} \and
      \inferrule{\Gamma \vdash t \inferty T \\ T \rtred \eq{A}{a}{a'}}{\Gamma \vdash t \pinferty{\eqind} \eq{A}{a}{a'}}
    \end{mathpar}

    \caption{Other bidirectional inductive types}
    \label{fig:bidir-indu-other}
  \end{figure}

  This same approach can be readily extended to other usual inductive types, with recursion or indices posing no specific problems, see \cref{fig:bidir-indu-other}. The choice to use $\pinferty{\uni}$ rather than $\checkty$ for types is guided by the intuition that the universe level of e.g.\ $A$ in $\eq{A}{a}{a'}$ is free, similarly to what happens for sum types.
  
  \begin{figure}
    \begin{mathpar}
      \jform{$\Gamma \vdash t \inferty T$}
      \inferrule{ }
      {\Gamma \vdash I\ulev{\lis{i}} \inferty \P (\lis{x} : \lis{X}\ulev{\lis{i}}) (\lis{y} : \lis{Y}\ulev{\lis{i}}), \uni[l\ulev{\lis{i}}]}[Ind] \label{infrule:indu-indu}\and
      \inferrule{ }
      {\Gamma \vdash c_k\ulev{\lis{i}} \inferty \P (\lis{x} : \lis{X}\ulev{\lis{i}}) (\lis{y} : \lis{Y_k}\ulev{\lis{i}}), I\ulev{\lis{i}}~\lis{x}~\lis{u_k\ulev{\lis{i}}} }[Cons] \label{infrule:indu-cons} \and
      \inferrule{
        \Gamma \vdash s \pinferty{I} I\ulev{\lis{i'}}~\lis{a}~\lis{b} \\
        \Gamma \vdash \lis{p}_k \checkty \subs{\lis{X}_k}{\lis{p}}{\lis{x}} \\
        \Gamma, \lis{y} : \subs{\lis{Y}\ulev{\lis{i}}}{\lis{p}}{\lis{x}}, z : I\ulev{\lis{i}}~\lis{p}~\lis{y} \vdash P \pinferty{\uni} \uni[j] \\
        I\ulev{\lis{i'}}~\lis{a}~\lis{b} \cumul I\ulev{\lis{i}}~\lis{p}~\lis{b} \\
        \Gamma, \lis{y} : \subs{\lis{Y_k}\ulev{\lis{i}}}{\lis{p}}{\lis{x}} \vdash \lis{t}_k \checkty \subs{\subs{P}{c_k\ulev{\lis{i}}~\lis{p}~\lis{y}}{z}}{\lis{u_k}\ulev{\lis{i}}}{\lis{y}}  
      }{\Gamma \vdash \match{s}{(I,\lis{i},\lis{p})}{P}{\lis{t}} \inferty \subs{\subs{P}{s}{z}}{\lis{b}}{\lis{y}}}[Match] \label{infrule:indu-match} \and
      \inferrule{
        \Gamma \vdash T \pinferty{\uni} \uni[i] \\
        \Gamma, f : T \vdash t \checkty T \\
        \text{ guard condition}
      }{\Gamma \vdash \fix{f}{T}{t} \inferty T}[Fix] \label{infrule:indu-fix} \\

      \jformlow{$\Gamma \vdash t \pinferty{I} T $}
        \inferrule{
          \Gamma \vdash t \inferty T \\
          T \red I~\lis{a}~\lis{b}
        }{\Gamma \vdash t \pinferty{I} I~\lis{a}~\lis{b}}

    \end{mathpar}

    \caption{Bidirectional inductive type – PCUIC style}
    \label{fig:bidir-indu-pcuic}
  \end{figure}

  \subparagraph{Polymorphic, Cumulative Inductive Types} 
  
  The account of general inductive types in PCUIC is quite different from the one we just gave. The main addition is universe polymorphism \cite{Sozeau2014}, which means that inductive types and constructors come with explicit universe levels. The $\S$-type of the previous paragraph, for instance, would contain an explicit universe level $i$, and both $A$ and $B$ would be checked against $\uni[i]$ rather than having their level inferred. This makes the treatment of complex inductive types possible by using checking uniformly – rather than relying on constrained inference to infer universe levels – at the cost of possibly needless annotations, as here with $\S$-types.
  To make that polymorphism more seamless, those polymorphic inductive types are also cumulative \cite{Timany2018}: in much the same way as $\uni[i] \cumul \uni[j]$ if $i \leq j$, also $\nat\ulev{i} \cumul \nat\ulev{j}$, where $\nat\ulev{i}$ denotes the polymorphic inductive $\nat$ at universe level $i$. This enables lifting from a lower universe to a higher one, so that for instance $\vdash 0\ulev{i} : \nat\ulev{j}$ if $i \leq j$.
  PCUIC as presented in MetaCoq also presents constructors and inductive types as functions, rather than requiring them to be fully applied, and it separates recursors into a pattern-matching and a fixpoint construct, the latter coming with a specific guard condition to keep the normalization property enjoyed by a system with recursors.

  A sketch of the bidirectional rules is given in \cref{fig:bidir-indu-pcuic}, for a generic inductive $I$.
  We use bold characters to denote lists – for instance $\lis{a}$ is a list of terms – and indexes to denote a specific element – so that $\lis{a}_k$ is the $k$-th element of the previous. The considered inductive $I$ has parameters of type $\lis{X}$, indices of type $\lis{Y}$ and inhabits some universe $\uni[l]$. Its constructors $c_k$ are of types $\P (\lis{x} : \lis{X}) (\lis{y} : \lis{Y_k}), I~\lis{x}~\lis{u}$, with $\lis{u_k}$ terms possibly depending on both $\lis{x}$ and $\lis{y}$. Since we are considering a polymorphic inductive type, all of those actually have to be instantiate with universe levels, an operation we denote with $\ulev{\lis{i}}$.
  
  The two rules \nameref{infrule:indu-indu}, \nameref{infrule:indu-cons} are similar to those for variables, with the types pulled out of a global environment – not represented in our rules – rather than of the context. In particular, this presentation completely fixes the universe levels of the arguments. In rule \nameref{infrule:indu-fix}, the type of the fixpoint is checked to be well-formed, and then the body is checked against it. The guard condition, although complex, does not vary between the directed and undirected systems, and we thus do not dwell on it. The formalised version of this rule is complicated by the need to consider mutual (co-)fixpoints, but follows the same pattern.
  
  Last but not least, \nameref{infrule:indu-match} follows the same structure as in \cref{fig:bidir-indu,fig:bidir-indu-other}: first, the type of the scrutinee is inferred, then the predicate is verified to be a type and finally the branches are checked.
  An important point is how much information can be retrieved from the scrutinee $s$. Indeed, the universe levels $\lis{i}$ and the parameters $\lis{p}$ used to build the context in which the predicate $P$ and branches $\lis{t}$ are typed are stored in the match constructor. For cumulative inductive types, this is crucial to retain equivalence between the undirected and bidirectional system, and wrongly building the context from the type inferred for the scrutinee led to the bug we discovered. The idea is that the match construction might need to be typed "higher" than the type of inferred for $s$ to be able to type $P$ and $\lis{t}$. Subsequently, a cumulativity check not appearing in the examples above is needed to ensure that the scrutinee checks against the type constructed using $\lis{i}$ and $\lis{p}$. In contrast with parameters, the inferred indices $\lis{b}$ can safely be used in the return type, but proving this is the most subtle part of the correctness proof.

  \subsection{The formalisation}

  Let us now go over the formalisation file by file.

  \subparagraph{\href{https://github.com/MevenBertrand/metacoq/blob/itp-artefact/bidirectional/theories/BDEnvironmentTyping.v}{BDEnvironmentTyping.v} \& \href{https://github.com/MevenBertrand/metacoq/blob/itp-artefact/bidirectional/theories/BDTyping.v}{BDTyping.v}}
  The first file refines a few definitions on contexts from \href{https://github.com/MevenBertrand/metacoq/blob/itp-artefact/template-coq/theories/EnvironmentTyping.v}{EnvironmentTyping.v} in order to account for the difference between checking and constrained inference. We expect this to eventually replace the less precise definitions.

  The second contains the definition of the bidirectional type system as a mutually defined inductive type whose main part is \href{https://github.com/MevenBertrand/metacoq/blob/2d631dcd91d2315e5a52fea0fdc27e59c30abd57/bidirectional/theories/BDTyping.v\#L29}{\coqe{infering}}.
  The best way to understand this inductive is probably to compare it with \href{https://github.com/MevenBertrand/metacoq/blob/2d631dcd91d2315e5a52fea0fdc27e59c30abd57/pcuic/theories/PCUICTyping.v\#L209}{\coqe{typing}}, the inductive predicate for undirected typing.

  We then go on to proving by hand a custom induction principle, by first introducing a notion of size of a derivation. This induction principle is not as strong as we might expect, as it does not provide the extra induction hypothesis on context and type that would go with McBride's discipline. We did not try to go and prove such a strong induction principle, as we did not need it. Instead, reflecting the discipline in the choice of the predicates proven by induction was enough. But the main reason was that proving such an induction principle effectively corresponds to an inline proof of validity, a property that required quite an important amount of work to get. We still conjecture that such a strong induction principle should be provable, by reproducing some of the lemmas on the undirected typing, with extra care taken to the size of the obtained typing derivations, so as to be able to use e.g.\ substitutivity of typing together with an induction hypothesis.

  \subparagraph{\href{https://github.com/MevenBertrand/metacoq/blob/itp-artefact/bidirectional/theories/BDToPCUIC.v}{BDToPCUIC.v} \& \href{https://github.com/MevenBertrand/metacoq/blob/itp-artefact/bidirectional/theories/BDFromPCUIC.v}{BDFromPCUIC.v}}
  The next two files prove the equivalence between both type system. Correctness (akin to \cref{thm:corr-ccomega}) is \href{https://github.com/MevenBertrand/metacoq/blob/2d631dcd91d2315e5a52fea0fdc27e59c30abd57/bidirectional/theories/BDToPCUIC.v\#L419}{\coqe{infering_typing}} for inference and the following theorems for the other judgements.
  Completeness (akin to \cref{thm:comp-cumul}) is theorem \href{https://github.com/MevenBertrand/metacoq/blob/2d631dcd91d2315e5a52fea0fdc27e59c30abd57/bidirectional/theories/BDFromPCUIC.v\#L387}{\coqe{typing_infering}}.

  The bulk of both proofs is an induction on typing derivation whose most challenging part is the handling of the case constructor, especially the subtle issues around indices described in \cref{sec:pcuic-indu}. Similarly to \cref{sec:ccomega}, correctness relies on the strong properties of validity and subject reduction to reconstruct missing premises, while completeness mostly requires transitivity of conversion and confluence of reduction.

  \subparagraph{\href{https://github.com/MevenBertrand/metacoq/blob/2d631dcd91d2315e5a52fea0fdc27e59c30abd57/bidirectional/theories/BDUnique.v}{BDUnique.v}} This last file contains the proofs of \cref{prop:unique-inf} – \href{https://github.com/MevenBertrand/metacoq/blob/2d631dcd91d2315e5a52fea0fdc27e59c30abd57/bidirectional/theories/BDUnique.v\#L347}{\coqe{uniqueness_inferred}} – and \cref{prop:princ-types} – \href{https://github.com/MevenBertrand/metacoq/blob/2d631dcd91d2315e5a52fea0fdc27e59c30abd57/bidirectional/theories/BDUnique.v\#L355}{\coqe{principal_type}}. Apart from some lemmas on conversion that were only proved for cumulativity in MetaCoq, the induction itself is quite straightforward thanks to the bidirectional structure.

\section{Beyond PCUIC}
  \label{sec:beyond}

  The use of our bidirectional structure is not limited to CIC or PCUIC. On the contrary, we found it crucial to have such a bidirectional type system when designing a gradual extension to CIC \cite{LennonBertrand2020}, for multiple reasons we try and detail below.
  
  But let us first give a bit of context about this extension. The aim was to adapt the ideas of gradual typing \cite{Siek2015} to CIC. Gradual typing aims at incorporating some level of dynamic typing into a static typing discipline. To do so, a new type constructor $\?$ is introduced to represent dynamic type information. At typing time, this $\?$ should be seen as a wildcard that represents "any type" that is to be treated optimistically. In particular, $\?$ should be considered convertible to any type. Conversion is thus replaced by a new relation, called consistency, that corresponds to this intuition. In effect it behaves somewhat similarly to unification, with each $\?$ corresponding to a unification variable. This means in particular that consistency is \emph{not} transitive, as if it were it would be useless: since any type $T$ is consistent with $\?$, if consistency were transitive any two types would be related.

\subparagraph{Localized computation}

  The free-standing conversion rule \nameref{infrule:pts-undir-conv} is powerful, but sometimes too much.

  This was our first use for the bidirectional structure. Indeed, multiple uses of a consistency in a row would have allowed to change the type of a term to any other arbitrary type by going through $\?$ using two conversion rules in a row. Thus, any term would have been typable! Being able to use the conversion rule unrestricted was too much. Instead, the bidirectional system enforces a more localized use of conversion: only once, at the interface between inference and checking. This restriction was enough to make the conversion rule meaningful again.

  More generally, since the equivalence between the undirected and directed variants relies on the properties listed in \cref{prop:ccomega}, when one of these fails the equivalence is endangered. When one envisons a system where this would be the case, the bidirectional approach might be worth considering, as it could stay viable while its undirected counterpart might not.

\subparagraph{Modes for the conversion rule}

  The observation made in \cref{sec:ccomega-bidir} that the unique \nameref{infrule:pts-undir-conv} rule serves two different roles, which is clarified by the separation between checking and constrained inference, is an important one when toying with computation. Indeed, those two different aspects must be accounted for if one wishes to modify conversion and/or reduction. In particular, modifying the definition of conversion without accounting for the specific role of reduction would make rules for checking and constrained inference come out of sync, bringing trouble down the road.
  
  Taking again the example of \cite{LennonBertrand2020}, the \nameref{infrule:pts-check} is modified by directly replacing conversion with the consistency relation usual in gradual typing. But this is not enough, because constrained inference must be handled as well. This is done by supplementing rule \nameref{infrule:pts-sort-inf} by another rule to treat the case when the inferred type reduces to the wildcard $\?$, that can be used as a type – with some care taken. The same happens for all constrained inference rules.

\subparagraph{Bidirectional elaboration}
  In works such as \cite{Saibi1997,Asperti2012,LennonBertrand2020}, the procedure described is not typing but rather elaboration: the subject of the derivation $t$ is in a kind of source syntax and the aim is not only to inspect $t$, but also to output a corresponding $t'$ in another target syntax. The term $t'$ is a more precise account of term $t$, for instance with solved meta-variables, inserted coercions, and so on. The bidirectional structure readily adapts to those settings, with the extra term $t'$ simply considered as an output of all judgements. As such, McBride's discipline as described in \cref{sec:ccomega-bidir} demands that when, in a context $\Gamma$, the subject $t$ elaborates to $t'$ while inferring type $T$, we should have $\Gamma \vdash t' : T$ – and similarly for all other typing judgements. Having all rules locally preserve this invariant ensures that elaborated terms are always well-typed.

\section{Related work}
  \label{sec:related}

\subsection{Constrained inference}
\label{sec:rel-constrained}

  Traces of constrained inference in diverse seemingly ad-hoc workarounds can be found in various works around typing for CIC, illustrating that this notion, although overlooked, is of interest.
  
  In \cite{Pollack1992}, $\Gamma \vdash t : T$ is used for what we write $\Gamma \vdash t \inferty T$, but another judgment written $\Gamma \vdash t :\geq T$ and denoting type inference followed by reduction is used to effectively inline the two hypothesis of our constrained inference rules. Checking is similarly inlined.

  Saïbi \cite{Saibi1997} describes an elaboration mechanism inserting coercions between types. Those are inserted primarily in checking, when both types are known. However he acknowledges the presence of two special classes to handle the need to cast a term to a sort or a function type without more informations, exactly in the places where we resort to constrained inference rather than checking.

  More recently, Sozeau \cite{Sozeau2007} describes a system where conversion is augmented to handle coercion between subset types. Again, $\Gamma \vdash t : T$ is used for inference, and the other judgments are inlined. Of interest is the fact that reduction is not enough to perform constrained inference, because type head constructors can be hidden by the subset construction: a term of subset type such as $\{f : \nat \to \nat \mid f~0 = 0 \}$ should be usable as a function of type $\nat \to \nat$. An erasure procedure is therefore required on top of reduction to remove subset types in the places where we use constrained inference.

  Abel and Coquand \cite{Abel2008} use a judgement written $\Delta \vdash V \delta \Uparrow \operatorname{Set} \rightsquigarrow i$, where a type $V$ is checked to be well-formed, but with its exact level $i$ free. This corresponds very closely to our use of $\pinferty{\uni}$.

  Traces can also be found in the description of Matita's elaboration algorithm \cite{Asperti2012}. Indeed, the presence of meta-variables on top of coercions makes it even clearer that specific treatment of what we identified as constrained inference is required. The authors introduce a special judgement they call type-level enforcing corresponding to our $\pinferty{\uni}$ judgement. As for $\pinferty{\Pi}$, they have two rules to apply a function, one where its inferred type reduces to a product, corresponding to \nameref{infrule:pts-prod-inf}, and another one to handle the case when the inferred type instead reduces to a meta-variable. As Saïbi, they also need a special case for coercions of terms in function and type position. However, their solution is different. They rely on unification, which is available in their setting, to introduce new meta-variables for the domain and codomain of a product type whenever needed. For $\pinferty{\uni}$ though this solution is not viable, as one would need a kind of universe meta-variable. Instead, they rely on backtracking to test multiple possible universe choices.

  Finally, we have already mentioned \cite{LennonBertrand2020} in \cref{sec:beyond}, where the bidirectional structure is crucial in describing a gradual extension to CIC. In particular, and similarly to what happens with meta-variables in \cite{Asperti2012}, all constrained inference rules are duplicated: there is one rule when the head constructor is the desired one, and a second one to handle the gradual wildcard.

\subsection{Completeness}

  Quite a few articles tackle the problem of bidirectional typing in a setting with an untyped – so called Curry-style – abstraction. This is the case of early work by Coquand \cite{Coquand1996}, the type system of Agda as described in \cite{Norell2007}, the systems considered by Abel in many of his papers \cite{Abel2007,Abel2008,Abel2011,Abel2017}, and much of the work of McBride \cite{McBride2016,McBride2018,McBride2019} on the topic. In such systems, $\lambda$-abstractions can only be checked against a given type, but cannot infer one, so that only terms with no $\beta$-redexes are typable. Norell \cite{Norell2007} argues that such $\beta$-redexes are uncommon in real-life programs, so that being unable to type them is not a strong limitation in practice. To circumvent this problem, McBride also adds the possibility of typing annotations to retain the typability of a term during reduction.
  
  While this approach is adapted to programming languages, where the emphasis is on lightweight syntax, it is not tenable for a proof assistant kernel, where all valid terms should be accepted. Indeed, debugging a proof that is rejected because the kernel fails to accept a perfectly well-typed term the user never wrote – as most proofs are generated rather than written directly – is simply not an option.

  In a setting with typed – Church-style – abstraction, if one wishes to give the possibility for seemingly untyped abstraction, another mechanism has to be resorted to, typically meta-variables. This is what is done in Matita \cite{Asperti2012}, where the authors combine a rule similar to 
  \nameref{infrule:pts-abs} – where the type of an abstraction is inferred – with another one, similar to the Curry-style one – where abstraction is checked – looking like this:
  \begin{mathpar} 
    \inferrule{T \rtred \P x : A' . B \\ \Gamma \vdash A \pinferty{\uni} \uni[i] \\ A \conv A' \\ \Gamma, x : A \vdash t \checkty B}{\Gamma \vdash \l x : A . t \checkty T}
  \end{mathpar}
  While such a rule would make a simple system such as that of \cref{sec:ccomega} “over-complete”, it is a useful addition to enable information from checking to be propagated upwards in the derivation. This is crucial in systems where completeness is lost, such as Matita's elaboration. Similar rules are described in \cite{Asperti2012} for let-bindings and constructors of inductive types.

  Although only few authors consider the problem of a complete bidirectional algorithm for type-checking dependent types, we are not the first to attack it. Already Pollack \cite{Pollack1992} does, and the completeness proof for \CCo{} of \cref{sec:ccomega} is very close to one given in his article. Another proof of completeness for a more complex CIC-like system can be found in \cite{Sozeau2007}. None of those however tackle as we do the whole complexity of PCUIC.
  
\subsection{Inputs and outputs}

  We already credited the discipline we adopt on well-formedness of inputs and outputs to McBride \cite{McBride2018,McBride2019}. A similar idea has also appeared independently in \cite{Bauer2020}. Bauer and his co-authors introduce the notions of a (weakly) presuppositive type theory \cite[Def.~5.6]{Bauer2020} and of well-presented premise-family and rule-boundary \cite[Def.~6.16 and 6.17]{Bauer2020} to describe a discipline similar to ours, using what they call the boundary of a judgment as the equivalent of our inputs and outputs. Due to their setting being undirected, this is however more restrictive, because they are not able to distinguish inputs from outputs and thus cannot relax their condition to only demand inputs to be well-formed but not outputs.

\section{Conclusion}

  We have described a judgmental presentation of the bidirectional structure of typing algorithms in the setting of dependent types. In particular, we identified a new family of judgements we called constrained inference. Those have no counterpart in the non-dependent setting, as they result from a choice of modes for the conversion rule, which is specific to the dependent setting. We proved our bidirectional presentation equivalent to an undirected one, both on paper on the simple case of \CCo{}, and formally in the much more complex and realistic setting of PCUIC. Finally, we gave various arguments for the usefulness of our presentation as a way to ease proofs, an intermediate between undirected type-systems and typing algorithms, a solid basis to design new type systems, and a tool to re-interpret previous work on type systems in a clearer way.

  Regarding future work, a type-checking algorithm is already part of MetaCoq, and we should be able to use our bidirectional type system to give a pleasant completeness proof by separating the concerns pertaining to bidirectionality from the algorithmic problems, such as implementation of an efficient conversion check or proof of termination.
  More broadly, bidirectional type systems should be an interesting tool in the feat of incorporating in proof assistants features that have been satisfactorily investigated on the theoretical level while keeping a complete and correct kernel, avoiding the pitfall of cumulative inductive type's incomplete implementation in Coq.
  A first step would be to investigate the discrepancies between the two kinds of presentations of inductive types \cref{sec:to-pcuic}, and in particular if all informations currently stored in the match node are really needed or if a more concise presentation can be given. But we could go further and study how to handle cubical type theory \cite{Vezzosi2019}, rewrite rules \cite{Cockx2021}, setoid type theory \cite{Altenkirch2019}, exceptional type theory \cite{Pedrot2018}, $\eta$-conversion…
  There might also be an interesting link to make with the current work on normalization by evaluation \cite{Abel2010} as an alternative to weak-head reduction for constrained inference.
  Finally, we hope that our methodology will be adapted as a base for other theoretical investigations. As a way to ease this adoption, studying it in a general setting such as that of \cite{Bauer2020} might be a strong argument for adoption.
%%
%% Bibliography
%%

%% Please use bibtex, 

\bibliography{biblio}

\end{document}